\documentclass[aip,jap,reprint,superscriptaddress,floatfix,longbibliography]{revtex4-2}

\usepackage{graphicx}
\usepackage{needspace}
\usepackage{amsmath}
\usepackage{textcomp}
\usepackage{siunitx}
\usepackage{pgfplots}
\pgfplotsset{compat=1.18,}
\usepgfplotslibrary{colormaps, groupplots}
\usepackage{tikz}
\usetikzlibrary{decorations.pathmorphing}
\tikzset{
    set/.style={
        execute at end picture={
            \coordinate (lower right) at (current bounding box.south east);
            \coordinate (upper left) at (current bounding box.north west);
        }
    },
    receive/.style={
        execute at end picture={
            \pgfresetboundingbox
            \path (upper left) rectangle (lower right);
        }
    }
}
\usepackage[hidelinks]{hyperref}

\hypersetup{pdftitle={Heat cleaning kinetics and thermal degradation of MOCVD-grown GaAs photocathodes},pdfauthor={G. Blume et al.}}

\begin{document}

\title{Heat cleaning kinetics and thermal degradation of MOCVD-grown GaAs photocathodes}

\author{G.~Blume}
\email{gtblume@jlab.org}
\affiliation{Department of Physics, Old Dominion University, Norfolk, VA 23529, USA}
\author{A.~Masters}
\affiliation{Department of Electrical and Computer Engineering, Old Dominion University, Norfolk, VA 23529, USA}
\author{C.~Kirk}
\affiliation{Department of Electrical and Computer Engineering, Old Dominion University, Norfolk, VA 23529, USA}
\author{S.~Poudel}
\affiliation{Department of Electrical and Computer Engineering, Old Dominion University, Norfolk, VA 23529, USA}
\author{J.~Hill}
\affiliation{Department of Electrical and Computer Engineering, Old Dominion University, Norfolk, VA 23529, USA}
\author{A.~Kachwala}
\affiliation{Thomas Jefferson National Accelerator Facility, Newport News, VA 23606, USA}
\author{J.~Grames}
\affiliation{Thomas Jefferson National Accelerator Facility, Newport News, VA 23606, USA}
\author{M.~L.~Stutzman}
\affiliation{Thomas Jefferson National Accelerator Facility, Newport News, VA 23606, USA}
\author{S.~Marsillac}
\affiliation{Department of Electrical and Computer Engineering, Old Dominion University, Norfolk, VA 23529, USA}
\author{M.~Grau}
\affiliation{Department of Physics, Old Dominion University, Norfolk, VA 23529, USA}

\date{\today}

\begin{abstract}
We evaluated metalorganic chemical vapor deposition (MOCVD)-grown GaAs/GaAsP strained-superlattice photocathodes with integrated distributed Bragg reflectors in a low-voltage Mott polarimeter at Jefferson Lab. 
Without a protective arsenic cap, surface contamination can develop during transport and installation. 
Heat cleaning removes this contamination by thermally desorbing material bound to the surface, but excessive heating can also damage the photocathode. 
We modeled quantum efficiency (QE) in terms of surface contaminant coverage to describe its increase with heating time. 
QE rose more rapidly and reached a plateau nearly twice as high at \SI{550}{\degreeCelsius} as at \SI{450}{\degreeCelsius}, while polarization remained near \SI{83}{\percent}. 
Further heating reduced QE and polarization. 
Reflectance measurements showed a shift of the optical resonance to shorter wavelengths, consistent with surface material loss. 
We also tested the cleaning method in a high-voltage photogun, where QE showed the same initial increase and subsequent decline. 
The cleaning procedure must therefore balance contaminant removal against loss of photocathode material.
\end{abstract}

\maketitle

\section{Introduction}

Spin-polarized accelerator programs rely on GaAs-based photocathodes to produce electron beams in DC high-voltage photoguns.\cite{adderley2024,grames2011} 
These programs require high electron-spin polarization and high quantum efficiency (QE).
Higher QE reduces the amount of laser power needed to produce a given photocurrent.\cite{liu2016} 
Strained thin-film photocathodes that demonstrated enhanced spin polarization\cite{maruyama1991,maruyama1992} led to the development of strained superlattices (SSL) photocathodes.
In SSL photocathodes, strain and quantum confinement lift the valence-band degeneracy to produce highly spin-polarized electrons.\cite{maruyama2004,kachwala2025} 
An integrated distributed Bragg reflector (DBR) beneath the SSL consists of layers with alternating refractive indices. 
It increases photon absorption by reflecting light back through the strained superlattice, increasing QE without a thicker absorbing layer and thereby reducing the laser power required
for a given beam current.\cite{liu2016}

Molecular beam epitaxy (MBE) has produced high-performance DBR SSL photocathodes,\cite{biswas2023} but limited production restricts their availability.\cite{belfore2023} 
Metalorganic chemical vapor deposition (MOCVD) offers higher growth rates and lower cost.\cite{belfore2023} 
Both methods have produced high-polarization SSL photocathodes.\cite{biswas2023,belfore2023,masters2026,grau2026cryogenic} 
MBE-grown surfaces can be protected during handling with an arsenic cap, which is removed before activation.\cite{bernstein1992,yang1992} 
The MOCVD-grown samples studied here had no protective arsenic cap. Exposure to air allows oxides and other contaminants to develop on the surface.\cite{biswas2020} 
Surface cleanliness is a determining factor in QE because it affects the quality of the negative-electron-affinity (NEA) surface formed during activation and the probability of electron escape.\cite{rodway1986,spicer1993}

Heat cleaning is an effective method to desorb surface contaminants from GaAs photocathodes before activation.\cite{biswas2020,blume2025preparation} 
Cleaning time and temperature govern contaminant removal, but excessive heating can also remove GaAs\cite{foxon1973} and degrade the photocathode. 
The photocathode surface and DBR form a Fabry--P\'erot optical cavity whose optical thickness determines the resonance wavelength.\cite{fabry1899,liu2016} 
Loss of surface material shortens the cavity and shifts this minimum to shorter wavelengths.
Previous measurements used optical resonance shifts to quantify GaAs sublimation, with scanning transmission electron microscopy confirming the material loss.\cite{blume2026sublimation}

This study examines how cleaning time and temperature affect performance at a fixed wavelength of \SI{780}{\nano\meter}, from the initial increase in QE to the degradation caused by further heating. 
A first-order thermal-desorption model describes the QE increase. 
Additional tests compare resonance changes after heating with those during charge extraction and evaluate the cleaning procedure in a high-voltage photogun.

\section{Samples and measurements}

We evaluated five MOCVD-grown samples that shared a common layer structure. 
The structure consists of a p-type GaAs substrate, a metamorphic GaAs$_{1-x}$P$_x$ buffer graded to $x=0.35$, a 12-pair DBR, a \SI{750}{\nano\meter} GaAs$_{0.65}$P$_{0.35}$ spacer, and a 14-pair SSL. 
The DBR alternates \SI{54}{\nano\meter} GaAs$_{0.65}$P$_{0.35}$ and \SI{64}{\nano\meter} In$_{0.30}$Al$_{0.70}$P layers. 
The SSL alternates \SI{3.8}{\nano\meter} compressively strained GaAs and \SI{2.8}{\nano\meter} GaAs$_{0.65}$P$_{0.35}$ layers, with a nominally \SI{5}{\nano\meter} highly doped GaAs surface layer. 
The buffer and DBR have p-type doping of $5\times10^{18}\,\mathrm{cm}^{-3}$, the SSL $5\times10^{17}\,\mathrm{cm}^{-3}$, and the surface layer $5\times10^{19}\,\mathrm{cm}^{-3}$.\cite{belfore2023,blume2026sublimation}

Before mounting each sample in vacuum, we etched it in \SI{20}{\percent} HCl for \SI{30}{\second}. 
We heated the photocathode with a rear-mounted heater rod. 
Because the heater and emitting surface are separated, we calibrated the heater setting against the surface temperature. 
To produce the NEA surface, we alternately supplied Cs and NF$_3$ under white-light illumination.\cite{liu2016b}
We repeated these ``yo-yo'' activation cycles until the gain in photocurrent was less than \SI{5}{\percent}.\cite{blume2025preparation}

For four samples, we measured QE and polarization in the retarding-field Mott polarimeter (miniMott) at Jefferson Lab.\cite{mccarter2010} 
Using a quarter-wave plate, we converted linearly polarized light from a SuperK supercontinuum source to circular polarization and reversed its helicity with an insertable half-wave plate. 
From photocurrent and incident optical power, we determined QE, the number of emitted electrons per incident photon.

To measure polarization, we accelerated the emitted electrons to \SI{20}{\kilo\electronvolt} and directed them onto a gold target. 
Detectors positioned at \SI{120}{\degree} to the incident beam counted the scattered electrons, determining the polarization through a rate asymmetry brought about by spin-dependent scattering off the target.\cite{sherman1956}
As inelastically scattered electrons dilute this asymmetry, we used retarding-field grids to reject electrons that had lost energy and linearly extrapolated the measured asymmetry to the elastic endpoint.
We obtained the polarization by dividing this asymmetry by the calibrated effective Sherman function, $0.201(5)$.\cite{mccarter2010}

We report QE and polarization at \SI{780}{\nano\meter}. 
This fixed illumination wavelength makes QE sensitive to changes in both emission probability and optical absorption. 
The insets in Fig.~\ref{fig:recovery} show an example wavelength scan. 
We measured the cavity-resonance wavelength $\lambda_R$ independently by reflectometry where the resonance appears as a minimum in the reflectance spectrum.\cite{blume2026sublimation}

We measured how QE and polarization changed with cumulative heating time at 450 and \SI{550}{\degreeCelsius} using sample~1 to identify when further heating produced little change in QE. 
This time series guided the choice of \SI{1.5}{\hour} treatments for samples~2 and 3, which we used to investigate the temperature response. 
The combined measurements span 450--\SI{580}{\degreeCelsius}. 
With a fourth sample, we compared resonance changes after heating with those during charge extraction. 
We tested the fifth sample in the DC high-voltage photogun at Jefferson Lab's Upgraded Injector Test Facility,\cite{blume2024} collecting photocurrent in a downstream Faraday cup and measuring polarization with a thin-foil Mott polarimeter.\cite{grames2020}

\section{Results and discussion}

\subsection{Dependence on heating time}
\label{sec:recovery}

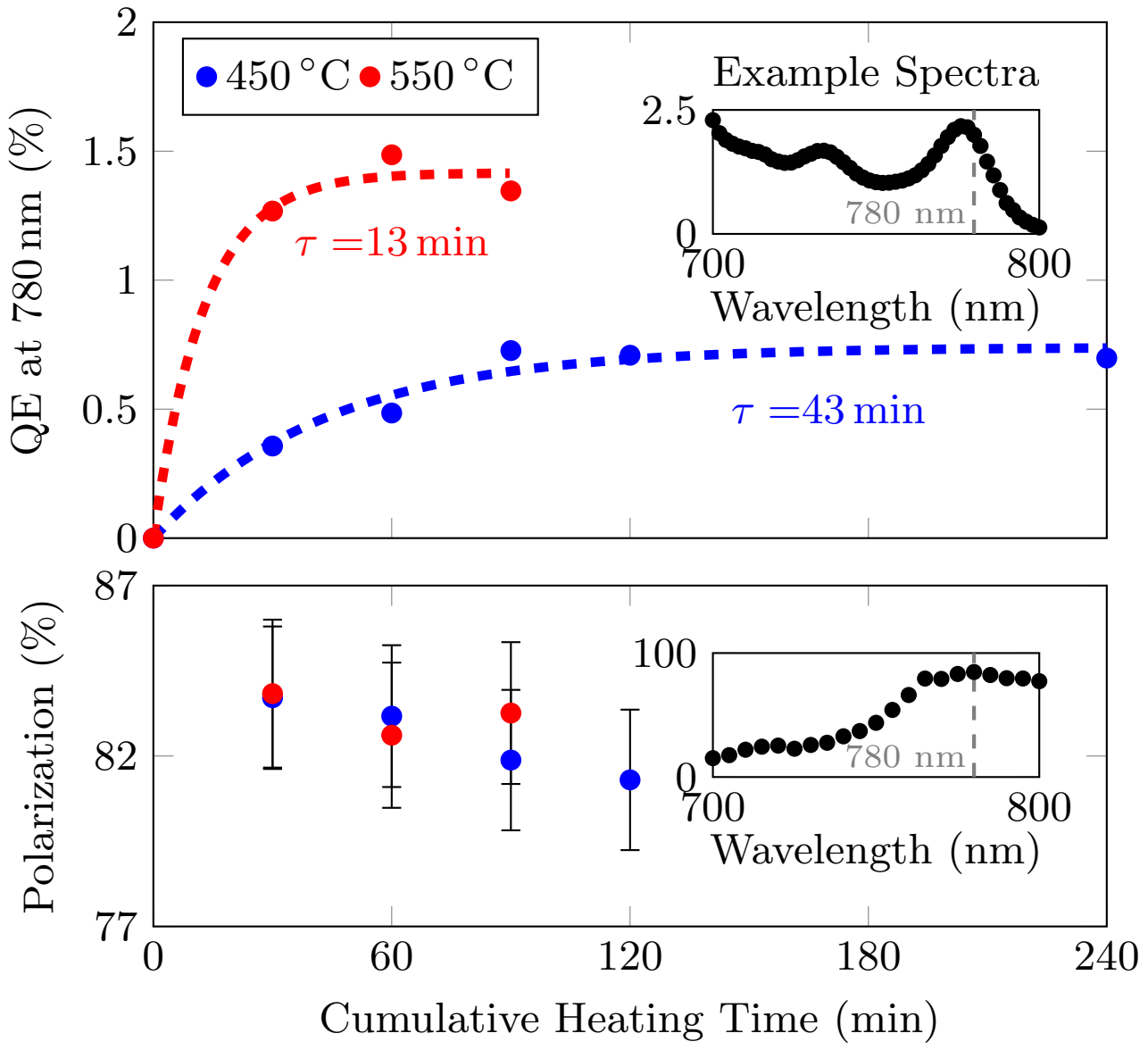
\begin{figure}[htbp!]
\begin{tikzpicture}[set]
\begin{groupplot}[
    group style={
        group name=myplots,
        group size=1 by 2,
        vertical sep=10pt
    },
    width=1\columnwidth,
    xmin=0, xmax=240
]
\nextgroupplot[
    height=.625\linewidth,
    xtick={0, 60, 120, 180, 240},
    xticklabel=\empty,
    ylabel={QE at \SI{780}{\nano\meter} (\SI{}{\percent})},
    ymin=0, ymax=2,
    legend style={font=\small},
    legend pos=north west, 
    legend columns=2
]
\addplot[only marks, blue] table[x, y index=1] {data/qe_time_data450.dat};
\addplot[only marks, red] table[x, y index=1] {data/qe_time_data550.dat};
\addplot[no marks, blue, dashed, line width=2pt, mark options={fill=black, draw=black}] table[x, y index=1] {data/fit450C.dat};
\node[blue] at (axis cs:170, 0.5) {$\tau=$\SI{43}{\minute}};
\addplot[no marks, red, dashed, line width=2pt, mark options={fill=black, draw=black}] table[x, y index=1] {data/fit550Cshort.dat};
\node[red] at (axis cs:60, 1.15) {$\tau=$\SI{13}{\minute}};
\legend{\SI{450}{\degreeCelsius},\SI{550}{\degreeCelsius}}

\nextgroupplot[
    height=.475\linewidth,
    ylabel={Polarization (\SI{}{\percent})},
    ymin=77, ymax=87,
    ytick={77, 82, 87},
    xtick={0, 60, 120, 180, 240},
    xlabel={Cumulative Heating Time (\SI{}{\min})},
    error bars/error bar style={black},
    error bars/y dir=both,
    error bars/y explicit, 
]
\addplot[only marks, blue] table[x=x, y=y, y error=err] {data/pol_time_data450.dat};
\addplot[only marks, red] table[x=x, y=y, y error=err] {data/pol_time_data550.dat};

\end{groupplot}

\begin{axis}[
    at={(myplots c1r1.north east)}, 
    anchor=north east, 
    width=4cm, 
    height=2.5cm, 
    xshift=-0.5cm,
    yshift=-0.65cm,
    ymin=0, ymax=2.5,
    xmin=700, xmax=800,
    ytick={0, 2.5},
    xtick={700, 800},
    title={Example Spectra},
    title style={yshift=-6pt},
    xlabel={Wavelength (\SI{}{\nano\meter})},
    xlabel style={yshift=4pt},
  ]
\addplot[mark size=1.5pt, black, only marks] table[x, y index=1] {data/dbr_qe_example.dat};
\draw[dashed, black!50!white, thick] (axis cs:780,0) -- (axis cs:780,2.5);
\node[black!50!white] at (axis cs:759, 0.425) {\scriptsize 780 nm};
\end{axis}

\begin{axis}[
    at={(myplots c1r2.north east)}, 
    anchor=north east, 
    width=4cm, 
    height=2.5cm, 
    xshift=-0.5cm,
    yshift=-0.5cm,
    ymin=0, ymax=100,
    xmin=700, xmax=800,
    ytick={0, 100},
    xtick={700, 800},
    error bars/error bar style={black},
    error bars/y dir=both,
    error bars/y explicit, 
    xlabel={Wavelength (\SI{}{\nano\meter})},
    xlabel style={yshift=4pt},
  ]
\addplot[mark size=1.5pt, black, only marks] table[x=x, y=y, y error=err] {data/dbr_polarization_example.dat};
\draw[dashed, black!50!white, thick] (axis cs:780,0) -- (axis cs:780,100);
\node[black!50!white] at (axis cs:759, 17) {\scriptsize 780 nm};
\end{axis}

\end{tikzpicture}
\caption{Cleaning-time response of sample~1 at \SI{780}{\nano\meter}. (a) QE and Eq.~(\ref{eq:recovery}). (b) Polarization, shown on the same time scale. Error bars show polarization uncertainties. Insets show example QE and polarization spectra. Dashed lines mark \SI{780}{\nano\meter}.}
\label{fig:recovery}
\end{figure}

At \SI{450}{\degreeCelsius}, QE increased over the first \SI{1.5}{\hour} and then remained near \SI{0.7}{\percent} through \SI{4}{\hour} of total heating [Fig.~\ref{fig:recovery}(a)]. 
At \SI{550}{\degreeCelsius}, QE rose more quickly and reached a higher plateau. 
We chose a first-order thermal-desorption model, which assumes that the rate is proportional to the remaining contaminant coverage.\cite{gunther2014} 
Assuming negligible QE before cleaning and an increase proportional to the fraction of the surface cleaned, we fitted
\begin{equation}
\eta(t,T)=\eta_f(T)\left[1-\exp(-t/\tau(T))\right],
\label{eq:recovery}
\end{equation}
where $\eta_f$ is the saturation QE and $\tau$ is the time constant for the approach to saturation.

The fits yield $\eta_f=\SI{0.74(2)}{\percent}$ and $\tau=\SI{43(4)}{\minute}$ at \SI{450}{\degreeCelsius}, compared with \SI{1.42(2)}{\percent} and \SI{13(1)}{\minute} at \SI{550}{\degreeCelsius}. 
Thus, heating at \SI{550}{\degreeCelsius} nearly doubled the saturation QE and reduced the time constant by a factor of 3.4. 
Polarization remained between approximately 81 and \SI{84}{\percent} over the shorter intervals shown in Fig.~\ref{fig:recovery}(b).

\begin{figure*}[htbp!]
\begin{tikzpicture}[set]
\begin{groupplot}[
    group style={
        group name=myplots,
        group size=3 by 1,
        horizontal sep=45pt
    },
    width=0.33\textwidth,
    height=0.33\linewidth,
    xmin=0, xmax=240
]
\nextgroupplot[
    xtick={420, 480, 540, 600},
    ytick={0, 0.5, 1, 1.5, 2, 2.5},
    ylabel={QE at \SI{780}{\nano\meter} (\SI{}{\percent})},
    ymin=0, ymax=2.5,
    xmin=420, xmax=600,
]
\addplot[mark size=1.5pt, red, only marks] table[x, y index=1] {data/heatstudysample2qedata.dat};
\addplot[mark size=1.5pt, blue, only marks] table[x, y index=1] {data/heatstudysample3qedata.dat};
\addplot[mark size=1.5pt, green!50!black, only marks] table[x, y index=1] {data/heatstudysample4qedata.dat};
\node[black] at (axis cs:580, 2.25) {\textbf{a)}};

\nextgroupplot[
    ylabel={Polarization at \SI{780}{\nano\meter} (\SI{}{\percent})},
    ymin=70, ymax=90,
    ytick={70, 75, 80, 85, 90},
    xtick={420, 480, 540, 600},
    xmin=420, xmax=600,
    xlabel={Cleaning Temperature (\SI{}{\degreeCelsius})},
    error bars/error bar style={black},
    error bars/y dir=both,
    error bars/y explicit,
    legend columns=3,
    legend style={
        at={(0.5, 1.2)},
        anchor=north
    }
]
\addplot[mark size=1.5pt, red, only marks] table[x=x, y=y, y error=err] {data/heatstudysample2poldata.dat};
\addplot[mark size=1.5pt, blue, only marks] table[x=x, y=y, y error=err] {data/heatstudysample3poldata.dat};
\addplot[mark size=1.5pt, green!50!black, only marks] table[x=x, y=y, y error=err] {data/heatstudysample4poldata.dat};
\node[black] at (axis cs:580, 88) {\textbf{b)}};
\legend{Sample 1, Sample 2, Sample 3}

\nextgroupplot[
    ylabel={Resonance $\lambda_R$ (\SI{}{\percent})},
    ymin=760, ymax=790,
    xtick={420, 480, 540, 600},
    xmin=420, xmax=600,
    ytick={760, 770, 780, 790},
    error bars/error bar style={black},
    error bars/y dir=both,
    error bars/y explicit, 
]
\addplot[only marks, red] table[x, y index=1] {data/heatstudysample2dbrdata.dat};
\addplot[only marks, blue] table[x, y index=1] {data/heatstudysample3dbrdata.dat};
\addplot[only marks, green!50!black] table[x, y index=1] {data/heatstudysample4dbrdata.dat};
\node[black] at (axis cs:580, 787) {\textbf{c)}};

\end{groupplot}

\end{tikzpicture}
\caption{Performance and optical response of samples~1--3: (a) QE at \SI{780}{\nano\meter}, (b) polarization at \SI{780}{\nano\meter}, and (c) cavity-resonance wavelength. Points appear at their measured temperatures without connecting lines. Samples~2 and 3 underwent successive \SI{1.5}{\hour} treatments. Sample~1's two comparison points accompany the time series in Fig.~\ref{fig:recovery}. Error bars show polarization uncertainties and \SI{1}{\nano\meter} wavelength uncertainties.}
\label{fig:temperature}
\end{figure*}
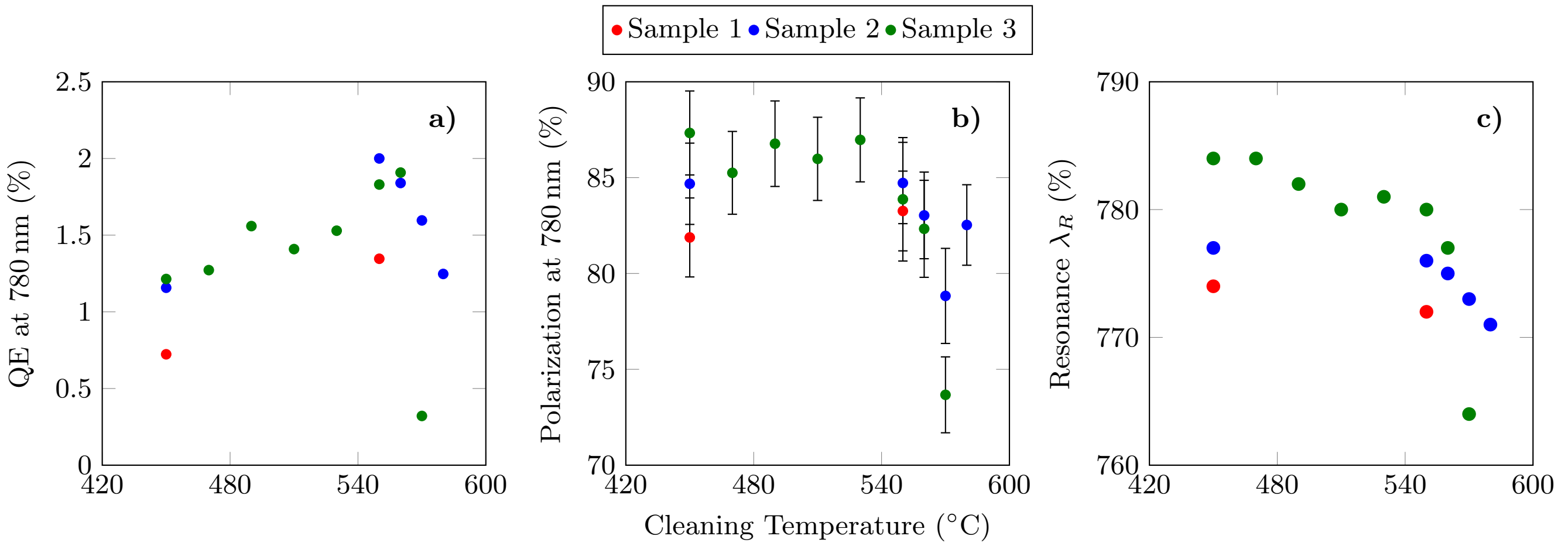

The higher QE plateau at \SI{550}{\degreeCelsius} is consistent with removal of contamination that remains after cleaning at \SI{450}{\degreeCelsius}. 
For example, Biswas et al.\cite{biswas2020} observed As$_2$O$_3$ removal from GaAs near \SI{500}{\degreeCelsius}, within our temperature range, whereas Ga$_2$O$_3$ required temperatures above \SI{600}{\degreeCelsius}. 
High temperatures can therefore improve QE, but heating enough to remove the remaining oxides also risks damage to the GaAs.

For thermally activated desorption, the rate constant $k(T)=1/\tau(T)$ follows the Arrhenius equation,
\begin{equation}
\frac{1}{\tau(T)}=\frac{1}{\tau_0}\exp\left(-E_a/(k_B T)\right),
\label{eq:desorption}
\end{equation}
where $\tau_0$ is a time prefactor, $E_a$ is the activation energy, and $k_B$ is Boltzmann's constant. 
The two time constants give an effective activation energy of approximately \SI{0.63}{\electronvolt} for the time dependent increase in QE.
The binding energy of H$_2$O on GaAs is reported to be \SI{0.7}{\electronvolt}\cite{MOKWA198498} suggesting the gradual improvement is consistent with the removal of water while the increase in saturation QE at higher temperature is consistent with the removal of surface oxides.

\subsection{Successive heat treatments}
\label{sec:temperature}

We heated samples~2 and 3 in successive \SI{1.5}{\hour} steps at increasing temperatures. 
Higher-temperature points therefore also correspond to longer total heating times. 
Figure~\ref{fig:temperature} compares all three samples. 
QE increased between 450 and \SI{550}{\degreeCelsius} for all three. 
Further heating reduced QE for sample~2, while sample~3 reached its highest measured QE at \SI{560}{\degreeCelsius} before a sharp decline at \SI{570}{\degreeCelsius}. 
Polarization also fell in the latter step, from approximately 82\,\% to 74\,\%.

The optical resonance measurements provide evidence of surface material loss during heating. 
Sample~3's resonance shifted from \SI{784}{\nano\meter} after its first treatment to \SI{764}{\nano\meter} after the \SI{570}{\degreeCelsius} treatment, when QE had fallen to approximately \SI{0.3}{\percent}. 
Sample~2 showed a smaller cumulative shift, from 777 to \SI{771}{\nano\meter} by \SI{580}{\degreeCelsius}. 
These shifts are consistent with loss of surface GaAs, which shortens the optical cavity.\cite{blume2026sublimation} 
The resulting change in resonant absorption contributes to the decrease in QE at the fixed \SI{780}{\nano\meter} illumination wavelength.

\begin{figure}[!htbp]
    \begin{tikzpicture}
    \node[inner sep=8pt] () at (0,0)
    {\includegraphics[width=0.45\textwidth]{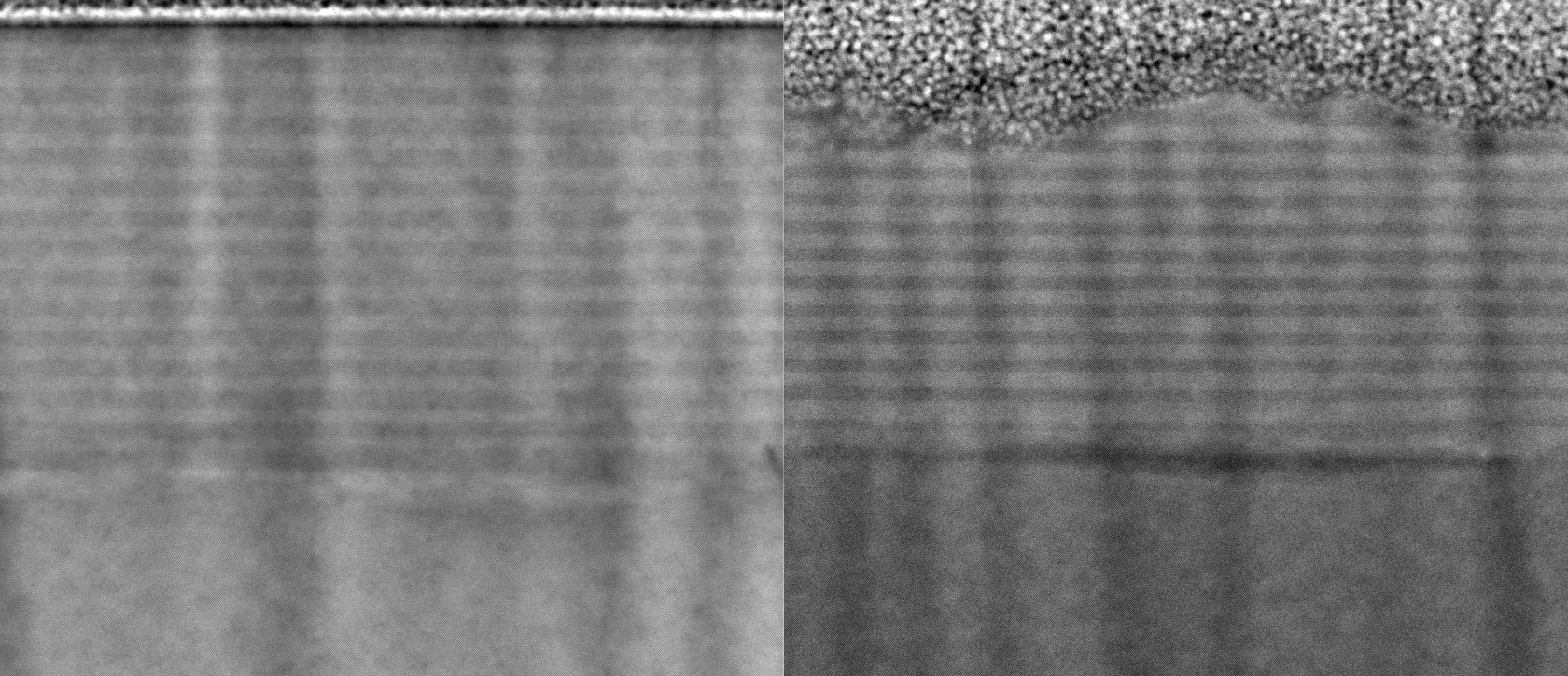}};
    \draw (-2.25,-1.2) node[anchor=north, align=center, yellow] {\textbf{No exposure}};
    \draw (2.25,-1.2) node[anchor=north, align=center, yellow] {\textbf{Post exposure}};
    \end{tikzpicture}
\caption{STEM comparisons (unscaled). Range spans roughly \SI{200}{\nano\meter}.}
\label{fig:tem}
\end{figure}

After the heat-treatment and charge-extraction measurements, we examined sample~2 by scanning transmission electron microscopy (STEM) and compared its surface and superlattice with an unexposed corner of the same wafer (Fig.~\ref{fig:tem}). 
The tested sample shows a rougher surface and loss of material. 
Thermal oxide desorption can also roughen GaAs surfaces.\cite{lee2006}

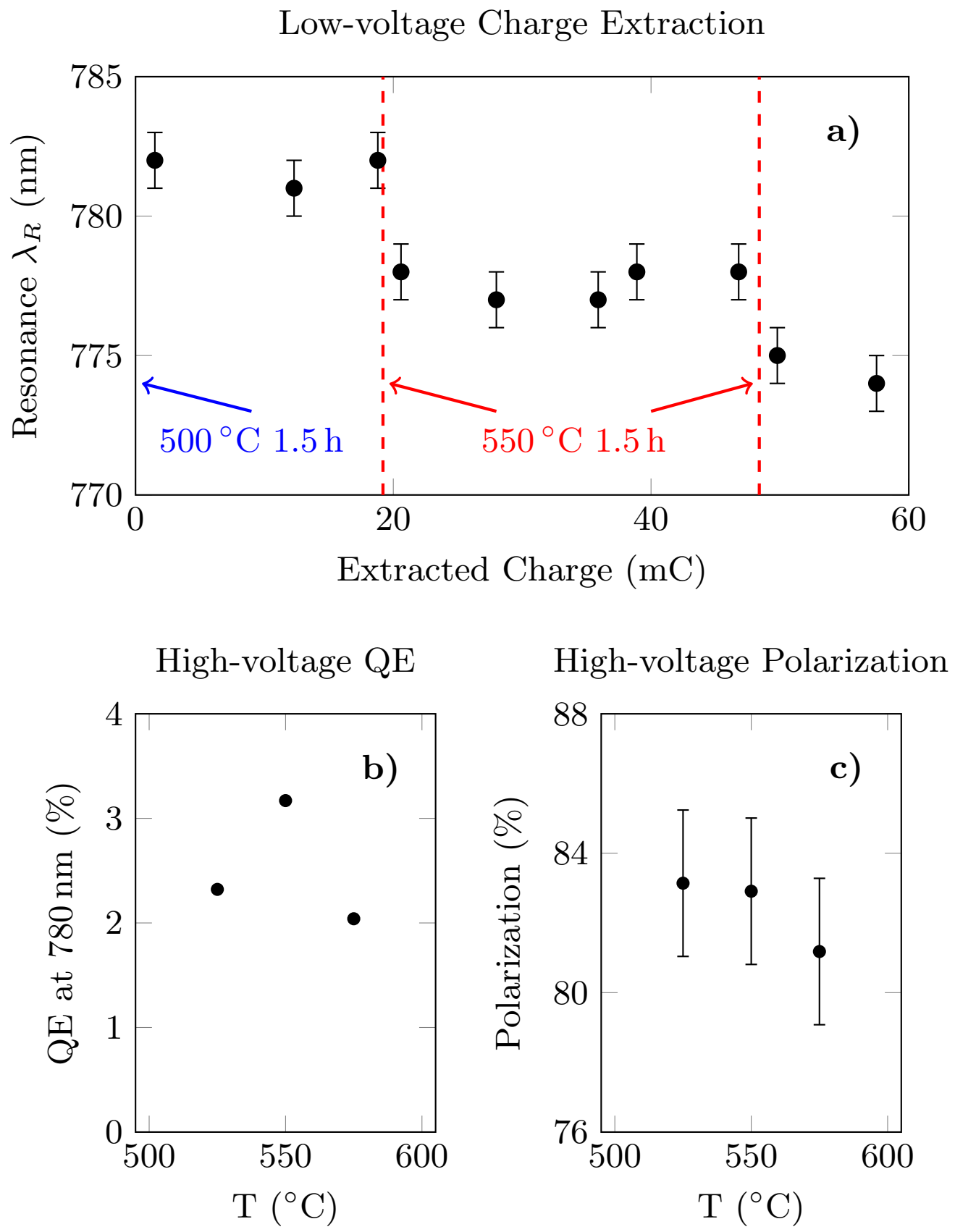
\begin{figure}[htbp!]
\begin{tikzpicture}[set]
\begin{groupplot}[
    group style={
        group size=2 by 2,
        vertical sep=2cm,
           horizontal sep=0.175\columnwidth
    },
    width=0.5\columnwidth, height=0.625\columnwidth
]    
\nextgroupplot[
    title=Top Plot, 
    width=\columnwidth, 
    xshift=0cm,
    height=.625\columnwidth,
    xmin=0, xmax=60,
    xtick={0, 20, 40, 60},
    title={Low-voltage Charge Extraction},
    xlabel={Extracted Charge (\SI{}{\milli\coulomb})},
    ylabel={Resonance $\lambda_R$ (\SI{}{\nano\meter})},
    ymin=770, ymax=785,
    ytick={770, 775, 780, 785},
    legend style={font=\small},
    legend pos=north west, 
    legend columns=2,
    error bars/error bar style={black},
    error bars/y dir=both,
    error bars/y explicit, 
]
\addplot[only marks, black] table[x=x, y=y, y error=err] {data/heatstudyiondamagestudy.dat};
\draw[dashed, red, thick] (axis cs:19.2,770) -- (axis cs:19.2,785);
\draw[dashed, red, thick] (axis cs:48.4,770) -- (axis cs:48.4,785);
\node[blue] at (axis cs:9, 772) {\SI{500}{\degreeCelsius} \SI{1.5}{\hour}};
\node[red] at (axis cs:34, 772) {\SI{550}{\degreeCelsius} \SI{1.5}{\hour}};
\draw[thick, blue, ->] (axis cs:9, 773) -- (axis cs:0.5, 774);
\draw[thick, red, ->] (axis cs:28, 773) -- (axis cs:19.7, 774);
\draw[thick, red, ->] (axis cs:40, 773) -- (axis cs:47.9, 774);
\node[black] at (axis cs:55, 783) {\textbf{a)}};
        
\nextgroupplot[group/empty plot,]

\nextgroupplot[
    title=High-voltage QE,
    xshift=-0.25\linewidth,
    xlabel={T (\SI{}{\degreeCelsius})},
    ylabel={QE at \SI{780}{\nano\meter} (\%)},
    ymin=0, ymax=4,
    xmin=495, xmax=605,
    ytick={0,1,2,3,4},
    xtick={500,550,600},
]
\addplot[mark size=1.5pt, black, only marks] table[x, y index=1] {data/uitfhvqe.dat};
\node[black] at (axis cs:585, 3.46) {\textbf{b)}};

\nextgroupplot[
    title=High-voltage Polarization, 
    xshift=-0.5\linewidth,
    xlabel={T (\SI{}{\degreeCelsius})},
    ylabel={Polarization (\%)},
    ymin=76, ymax=88,
    xmin=495, xmax=605,
    ytick={76,80,84,88},
    xtick={500,550,600},
    error bars/error bar style={black},
    error bars/y dir=both,
    error bars/y explicit, 
]
\addplot[mark size=1.5pt, black, only marks] table[x=x, y=y, y error=err] {data/uitfhvpol.dat};
\node[black] at (axis cs:585, 86.4) {\textbf{c)}};

\end{groupplot}

\end{tikzpicture}
\caption{Charge-extraction and photogun measurements. (a) Cavity resonance versus extracted charge for sample~4. Dashed lines mark \SI{1.5}{\hour} heat treatments: \SI{500}{\degreeCelsius} before charge extraction and two subsequent treatments at \SI{550}{\degreeCelsius}. Wavelength uncertainties are \SI{1}{\nano\meter}. (b) QE and (c) polarization at \SI{780}{\nano\meter} for sample~5 in the high-voltage photogun, after successive \SI{1.5}{\hour} treatments at 525, 550, and \SI{575}{\degreeCelsius}. Error bars in (c) show polarization uncertainties.}
\label{fig:controls}
\end{figure}

\subsection{Charge extraction and photogun measurements}
\label{sec:controls}

The damage and loss of surface material could arise from heating or from ion back-bombardment during charge extraction.\cite{grames2011,shutthanandan2012,yoskowitz2024} 
To compare resonance changes associated with heating and charge extraction, we heated a fourth sample to \SI{500}{\degreeCelsius} for \SI{1.5}{\hour} and then extracted approximately \SI{20}{\milli\coulomb}. 
We then extracted charge after each of two further \SI{1.5}{\hour} treatments at \SI{550}{\degreeCelsius}, reaching approximately \SI{60}{\milli\coulomb} in total. 
The resonance wavelength shifted after each additional heat treatment but showed no systematic change during the intervening charge extraction [Fig.~\ref{fig:controls}(a)].

To determine whether the temperature effects also occurred at high voltage, we tested the fifth sample in the photogun after successive \SI{1.5}{\hour} treatments at 525, 550, and \SI{575}{\degreeCelsius}. 
QE increased from approximately \SI{2.3}{\percent} to \SI{3.4}{\percent} after heating to \SI{550}{\degreeCelsius} and then fell to \SI{2.0}{\percent} at \SI{575}{\degreeCelsius} [Fig.~\ref{fig:controls}(b)]. 
This reproduces the rise and subsequent fall observed in the low-voltage measurements. 
Polarization remained near \SI{83}{\percent} for the first two treatments and reached about \SI{81}{\percent} for the last [Fig.~\ref{fig:controls}(c)].

The faster cleaning at \SI{550}{\degreeCelsius} favors limiting high-temperature exposure once QE approaches its plateau. 
The measurements motivate a shorter high-temperature step following the initial \SI{450}{\degreeCelsius} cleaning.

\section{Conclusion}

Heat cleaning improves QE in uncapped MOCVD-grown GaAs photocathodes, but prolonged exposure to temperatures above \SI{500}{\degreeCelsius} will damage the surface. 
Raising the temperature from 450 to \SI{550}{\degreeCelsius} nearly doubled the saturation QE and reduced the time constant by about a factor of three, while polarization remained approximately constant. 
Further heating reduced QE and polarization, with optical resonance shifts indicating loss of surface material. 
These results favor limiting high-temperature exposure once QE approaches its plateau.

\section*{Acknowledgments}

This material is based upon work supported by the U.S. Department of Energy, Office of Nuclear Physics, Contract Number DE-SC0025519.
This material is based upon work supported by the U.S. Department of Energy, Office of Science, Office of Nuclear Physics under Contract No. 89243126CSC000213.
This work was performed in part at the Analytical Instrumentation Facility (AIF) at North Carolina State University, which is supported by the State of North Carolina and the National Science Foundation (award number ECCS-2025064). 
The AIF is a member of the North Carolina Research Triangle Nanotechnology Network (RTNN), a site in the National Nanotechnology Coordinated Infrastructure (NNCI).

\par\vspace{12pt}
\noindent\begin{minipage}{\columnwidth}
\section*{Author declarations}
\subsection*{Conflict of interest}
The authors have no conflicts of interest to disclose.
\subsection*{Author contributions}
TBD
\end{minipage}

\par\vspace{12pt}
\noindent\begin{minipage}{\columnwidth}
\section*{Data availability}
The corresponding author will provide the data supporting this study on reasonable request.
\end{minipage}

\bibliography{references}
\end{document}